\documentclass{ws-procs975x65}

\usepackage{graphicx,epsfig, amsmath,amssymb,amsfonts}

\begin{document}


\title{CONSTRAINTS ON DARK ENERGY EQUATION OF STATE PARAMETERS FROM COSMIC TOPOLOGY}

\author{M. J. REBOU\c{C}AS and A. F. F. TEIXEIRA}
\address{Centro Brasileiro de Pesquisas F\'{\i}sicas\\
Rua Dr.\ Xavier Sigaud 150, \  22290-180 Rio de Janeiro --
RJ, Brazil}

\vspace{-3mm}
\begin{abstract}
Despite our present-day inability to predict the topology of the universe it is
expected that we should be able to detect it in the near future. A nontrivial
detectable topology of the space section of the universe can be probed for all
homogeneous and isotropic universes through the circles-in-the-sky. We discuss
briefly how one can use this observable attribute to set constraints
on the dark energy equation of state parameters.
\end{abstract}

\bodymatter
\vspace{-3mm}
\section{Introduction}\label{intro}

In standard  cosmology the Universe is described by
a $4$--manifold $\mathcal{M} = \mathbb{R}\times M_3$
endowed with the spatially homogeneous and isotropic
Friedmann-Lema\^{\i}tre-Robertson-Walker (FLRW) metric.
Furthermore, the spatial sections $M$ are often assumed
to be the simply-connected $3$--manifolds:
Euclidean $\mathbb{E}^3$, spherical $\mathbb{S}^3$, or
hyperbolic space $\mathbb{H}^3$.
However, the majority of constant curvature $3$--spaces, $M_3$,  are
multiply-connected quotient manifolds of the form $\mathbb{R}^3/\Gamma$,
$\mathbb{S}^3/\Gamma$, and $\mathbb{H}^3/\Gamma$, where $\Gamma$ is
fixed-point free group of isometries (see, e.g., the review
Refs.~\refcite{CosmTopReviews} for details).
An important observational consequence of a nontrivial (multiply-connected)
observable spatial non-trivial topology\cite{TopDetec} of $M_3$ is the
existence of the circles-in-the-sky,\cite{CSS1998} i.e.
pairs of matching circles with the same distribution of
temperature fluctuations, identified by $\Gamma$.
Hence, to observationally probe a putative non-trivial topology of $M_3$,
one ought to extract the pairs of correlated circles from full-sky CMBR maps.

Here we complement concisely our previous works\cite{Previous}
by showing how a possible detection of a pair of circles-in-the-sky
can be used to set constraints on the dark energy equation
of state (EOS) parameters in \emph{globally homogeneous} universes.

\vspace{-0.3cm}
\section{Topological  Constraints on EOS parameters}
\label{MainRes}

\begin{figure}[!htb]
\begin{center} \vspace{-0.5cm}
\includegraphics[width=4.5cm,height=4cm,angle=0]{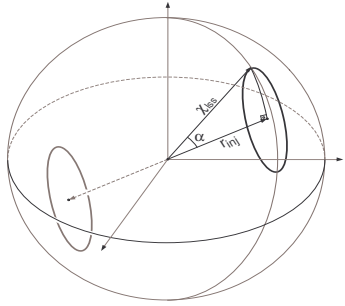}
\caption{A pair of antipodal circle in globally homogeneous
universes with a nontrivial topology. \label{Fig1} }
\vspace{-0.6cm}
\end{center}
\end{figure}

To show  how a possible detection of a nontrivial spatial topology
of globally homogeneous universes can be used to place constraints
on the dark energy equation of state, we begin by recalling that
in these spaces the pairs of correlated circles are
antipodal, as shown in Figure~\ref{Fig1}.\footnote{Recent searches
restricted to antipodal or nearly antipodal (deviation from antipodicity
$\theta \le 10^{\circ}$) circles and with radii $\alpha \gtrsim 18^{\circ}$
have been undertaken without success~\cite{Cornish-et-al-03,Key-et-al-07}.
Thus, there is a range of radii ($\alpha \lesssim 18^{\circ}$) that was not
covered by these searches. See also Refs.~\refcite{RelatedCirc} for
discussions about the negative outcome of these searches.}

Since there are no Clifford translations in the hyperbolic
geometry, there are no globally homogeneous hyperbolic manifolds.
Thus, for the sake of brevity we focus on  the globally homogeneous
spherical manifolds (Ref.~\refcite{Reboucas2009} contains a
discussion of the flat case)
In this case a straightforward use of trigonometric
relations for the right-angled spherical triangle shown in
Fig.~\ref{Fig1} yields
\begin{equation} \label{Obs_ChiLSS}
\chi^{\rm obs}_{lss} = \tan^{-1} \left[\,\frac{\tan r_{inj}}{
\cos \alpha}\, \right] \,,
\end{equation}
where $r_{inj}$ is a topological invariant, whose values are known for
all globally homogeneous spherical spaces (see Table~I of
Ref.~\refcite{Reboucas2009}), and the distance $\chi^{\rm obs}_{lss}$
is the comoving distance to the last scattering surface (LSS)
\emph{in units of the present-day curvature radius},
$a_0=a(t_0)=(\,H_0\sqrt{|1-\Omega_{\mathrm{tot}}|}\,)^{-1}\,$
($k \neq 0$).

Now, a circles-in-the-sky detection means a measurement of a value
for the radius $\alpha$ with an unavoidable observational
uncertainty $\sigma_\alpha$. These observational data along
with Eq.~(\ref{Obs_ChiLSS}) and the usual error propagation
formula, give the observational distance $\chi^{\rm obs}_{lss}$
to the LSS and the associated uncertainty $\sigma_{\chi_{lss}^{}}$.

On the theoretical side, the comoving distance to the last scattering
surface in units of the curvature radius is given by
\begin{equation}
\label{ChiLSS}
\chi_{lss}^{\rm th} = \frac{d^{}_{lss}}{a_0} = \sqrt{|\Omega_k|}\,
\int_1^{1+z_{lss}} \, \frac{H_0}{H(x)} \,\, dx \,,
\end{equation}
where $d^{}_{lss}$ is the radius of the LSS,
$x=1+z$ is an integration variable, $H$ is the Hubble
parameter, $\Omega_k =1-\Omega_{\mathrm{tot}}$ is the curvature density
parameter, and $z_{lss} \simeq 1089$. Clearly, different
parametrizations of the equation of state $\omega_x=p_x / \rho_x$ give
rise to different Friedmann equations, i.e.,  different ratios $H(z)/H_0\,$.
As an example, assuming that the current matter content of the Universe
is well approximated by a dust of density $\rho_m$ (baryonic plus dark matter)
along with a dark energy perfect fluid component of density $\rho_x$
and pressure $p_x$, for the parametrization $\omega_x= \omega_0 + \omega_1\, z /(1+z)$
(see Refs.~\refcite{Chevallier2001-Linder2003})
the Friedmann equation takes the form
\begin{eqnarray}
\left({ \frac{H}{H_0} }\right)^2 &=& \Omega_{m0}(1+z)^{3}+\Omega_{k0}(1+z)^{2}
+\Omega_{x0}(1+z)^{3(\omega_0+\omega_1 + 1)} \exp (- \frac {3\omega_1 z}{1+z})
\label{CPL} \,.
\end{eqnarray}

By comparing the observational topological value $\chi^{\rm obs}_{lss}$
for a given topology  with the theoretical values $\chi_{lss}^{\rm th}$ for
any dark energy equation of state parametrization one can set constraints on
the dark energy equation of state parameters.
In this way, the constraints from a detectable spatial topology are
taken into account in a $\chi^2$ statistical  analysis of any parameter
plane (as $\omega_1-\,\Omega_k\,$ and $\omega_0-\,\omega_1$, for example)
by adding  a term of the form
\begin{equation} \label{chisqtop}
\chi^2_{\rm top}= \left(\frac{\chi^{\rm obs}_{lss} -\chi_{lss}^{\rm th}}%
{\sigma_{\chi_{lss}}}\right)^2
\end{equation}
to the remaining $\chi^2$ terms that account for other
observational data sets. A concrete application of this procedure can
be found in Ref.~\refcite{SLR}.

\vspace{-4mm}

\section*{Acknowledgments}

This work is supported by Conselho Nacional de Desenvolvimento
Cient\'{\i}fico e Tecnol\'{o}gico (CNPq) - Brasil, under grant
No. 472436/2007-4. M. J. Rebou\c{c}as thanks CNPq for the grant
under which this work was carried out.


\end{document}